\documentclass[runningheads]{llncs}
\usepackage[T1]{fontenc}
\usepackage{graphicx}

\usepackage{subcaption}
\usepackage{cite}
\usepackage{footnote}

\usepackage{amssymb}
\usepackage{makecell}
\usepackage{booktabs}
\usepackage[svgnames,table]{xcolor}
\newcommand{\rf}[1]{\cellcolor{black!20}{#1}}
\renewcommand{\bf}[1]{\cellcolor{black!20}\textbf{#1}}

\usepackage{url}
\usepackage[hidelinks]{hyperref}
\usepackage{academicons}
\newcommand{\orcid}[1]{\href{https://orcid.org/#1}{\textcolor[HTML]{A6CE39}{\includegraphics[scale=0.5]{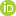}}}}

\begin{document}
\title{Multi-Modal Recommendation System with Auxiliary Information}
%
%
\author{Mufhumudzi Muthivhi\inst{1}\orcid{0000-0003-0509-6235} \and
Terence L. van Zyl\inst{2}\orcid{0000-0003-4281-630X} \and
Hairong Wang\inst{1}\orcid{0000-0001-8770-5916}}
\authorrunning{M. Muthivhi et al.}
%
\institute{University of the Witwatersrand, Johannesburg GT 2000, South Africa 
\email{1599695@students.wits.ac.za, hairongwng@gmail.com} \and
University of Johannesburg, Johannesburg GT, 2092, South Africa
\email{tvanzyl@uj.ac.za}}
\maketitle              
\begin{abstract}
Context-aware recommendation systems improve upon classical recommender systems by including, in the modelling, a user's behaviour. Research into context-aware recommendation systems has previously only considered the sequential ordering of items as contextual information. However, there is a wealth of unexploited additional multi-modal information available in auxiliary knowledge related to items.
This study extends the existing research by evaluating a multi-modal recommendation system that exploits the inclusion of comprehensive auxiliary knowledge related to an item.
The empirical results explore extracting vector representations (embeddings) from unstructured and structured data using data2vec. The fused embeddings are then used to train several state-of-the-art transformer architectures for sequential user-item representations.
The analysis of the experimental results shows a statistically significant improvement in prediction accuracy, which confirms the effectiveness of including auxiliary information in a context-aware recommendation system. We report a 4\% and 11\% increase in the NDCG score for long and short user sequence datasets, respectively.

\end{abstract}
\section{Introduction}
Web 2.0 gave rise to an information overload. In response, the industry has introduced Recommendation Systems to suggest a relevant item to users based on their preferences~\cite{billsus1998learning}. 
Recommending the most relevant item to a user requires an understanding of item characteristics and user preferences~\cite{adomavicius2005toward}.
Recently, researchers have proposed sequential recommendation to predict the next item that a user would prefer. Sequential recommendation models the historical consumption of items made by a user. Some state-of-the-art recommendation models only use item identifiers to model user behaviour~\cite{he2017translation, kang2018self, sun2019bert4rec}. Item identifiers are unique numbers assigned to each item. Furthermore, by exploiting the various forms of auxiliary information, we can enhance the modelling of user behaviour~\cite{zhang2019deep}.

One area of research has explored the use of auxiliary information through tabular data. They consider the keyword descriptions of items. Keyword descriptions are attribute data in tabular form usually transformed into one-hot encoded embeddings~\cite{fischer2020integrating, vasile2016meta, arunmetatransformer4rec}. However, their work neglects the use of unstructured data for the sequential recommendation task. This study explores an enhanced user-item representation learning framework. We use multi-modal auxiliary information whilst evaluating state-of-the-art transformer architectures that exploit the sequential dependencies between items.


The results demonstrate that transformer architectures with multi-modal auxiliary information improved the prediction results. Furthermore, SASRec, a unidirectional transformer architecture, displayed improvement over the bidirectional implementations. Finally, our analysis shows that the presented model considers each user's context and can make recommendations of items across the different categories.

We contribute to the existing literature by:
\begin{itemize}
    \item using multi-modal auxiliary information, including both text and image data to enhance the sequential recommendation task;
    \item using an embedding of continuous values instead of a multi-hot encoding of categorical data.
    \item using XGBoost to extract embeddings from tabular data
    \item showing that a unidirectional transformer, SASRec, yields the greatest improvement over the bidirectional transformers, BERT4Rec; and
    \item conducting an ablation study to analyse the effects of the various forms of modalities and their combinations.
\end{itemize}

\section{Background and Related Work}

Recommendation datasets consist of user feedback data. User feedback data provides information on whether a user has consumed an item or not. It may come in the form of a rating, purchase or webpage view of an item by a user. Users rate a small percentage of the item collection. Researchers refer to the existence of finite ratings between user-item pairs as the \textit{data sparsity} problem~\cite{sarwar2000application}. As a result, recommendation models have to predict the next item a user will consume from limited historical data.

Early work employed Matrix Factorization (MF) methods to find the latent factor space that encodes the user-item interaction matrix~\cite{koren2009matrix}. Koren~\textit{et al.} found that a combination of implicit and explicit feedback alleviates the data sparsity problem and increases prediction accuracy~\cite{koren2010factor}. It became increasingly evident that understanding item characteristics and user preferences would aid in recommending the most relevant item to a user~\cite{adomavicius2005toward}. Koren's findings~\cite{koren2010factor} motivated researchers to expand their experiments towards auxiliary information in the form of keyword descriptions.

Subsequently, research in recommendation systems focused on finding a recommendation dataset's latent factor space. Hence, recommendation systems have predominately become a representation learning problem. The representation learning task led to using neural networks, which learn non-linear patterns from user-item data~\cite{zhang2019deep}.
Simultaneously, within computer vision and natural language processing (NLP), researchers found that effective data representations significantly improve existing models~\cite{devlin2018bert, hjelm2018learning, kong2019mutual}.
Inspired by the developments in NLP, computer vision and other fields, the authors believe that the relevance of an item recommended to a user depends on the quality of an item's representation~\cite{dlamini2019author,van2020unique,manack2020deep,variawa2022transfer}. By exploiting comprehensive knowledge from item and user auxiliary information, we can design a rich contextualized embedding space of users, items, and their relationships.
Auxiliary knowledge exists in a variety of formats. Item descriptions and user profile information may be tabular or text/image/audio format (unstructured). In addition, we may also have the time at which an item is consumed, known as sequential data. The combination of these various forms of data allows us to model user behaviour effectively~\cite{zhang2019deep}.


Researchers have employed auxiliary information in recommendation systems to understand user behaviour. They model the sequential patterns within a user's consumption history and make predictions of the following preferred item. Recently, several novel transformer architectures have emerged for the sequential recommendation task. Kang~\textit{et el.} proposed a two-layer Transformer decoder that models users' sequential behaviour, known as SASRec~\cite{kang2018self}. However, Sun~\textit{et el.} argue that unidirectional models, like SASRec, do not sufficiently learn optimal representations for user behaviour sequences~\cite{devlin2018bert}. Inspired by the success of BERT~\cite{devlin2018bert}, the authors proposed a bidirectional model called BERT4Rec~\cite{sun2019bert4rec}. Then, Fischer~\textit{et al.} improved on BERT4Rec by integrating auxiliary information in the form of keyword embedding vectors~\cite{fischer2020integrating}.

The current trend in research is to append keyword descriptions of items to item identifiers~\cite{fischer2020integrating, zhou2020s3}. Keyword descriptions are attribute data in tabular form. In most cases, the data is categorical and transformed into a one-hot encoded embedding.
This paper argues that comprehensive knowledge about users and items exists in different modalities. Extracting this knowledge should result in better embeddings of users and items. We aim to reflect on the significance of auxiliary knowledge by exploiting various forms of data, including tabular, unstructured and sequential data.

We contribute to the field by exploring the design of an enhanced modelling process of user behaviour. We explore extracting embeddings from tabular data using gradient boosting and from unstructured data using data2vec. We fuse both structured and unstructured embeddings and learn the sequential behaviour of a user's preference through a unidirectional and bidirectional transformer architecture, SASRec and BERT4Rec respectively~\cite{kang2018self,sun2019bert4rec}

\begin{figure}[h]
\centerline{\includegraphics[width=0.8\columnwidth]{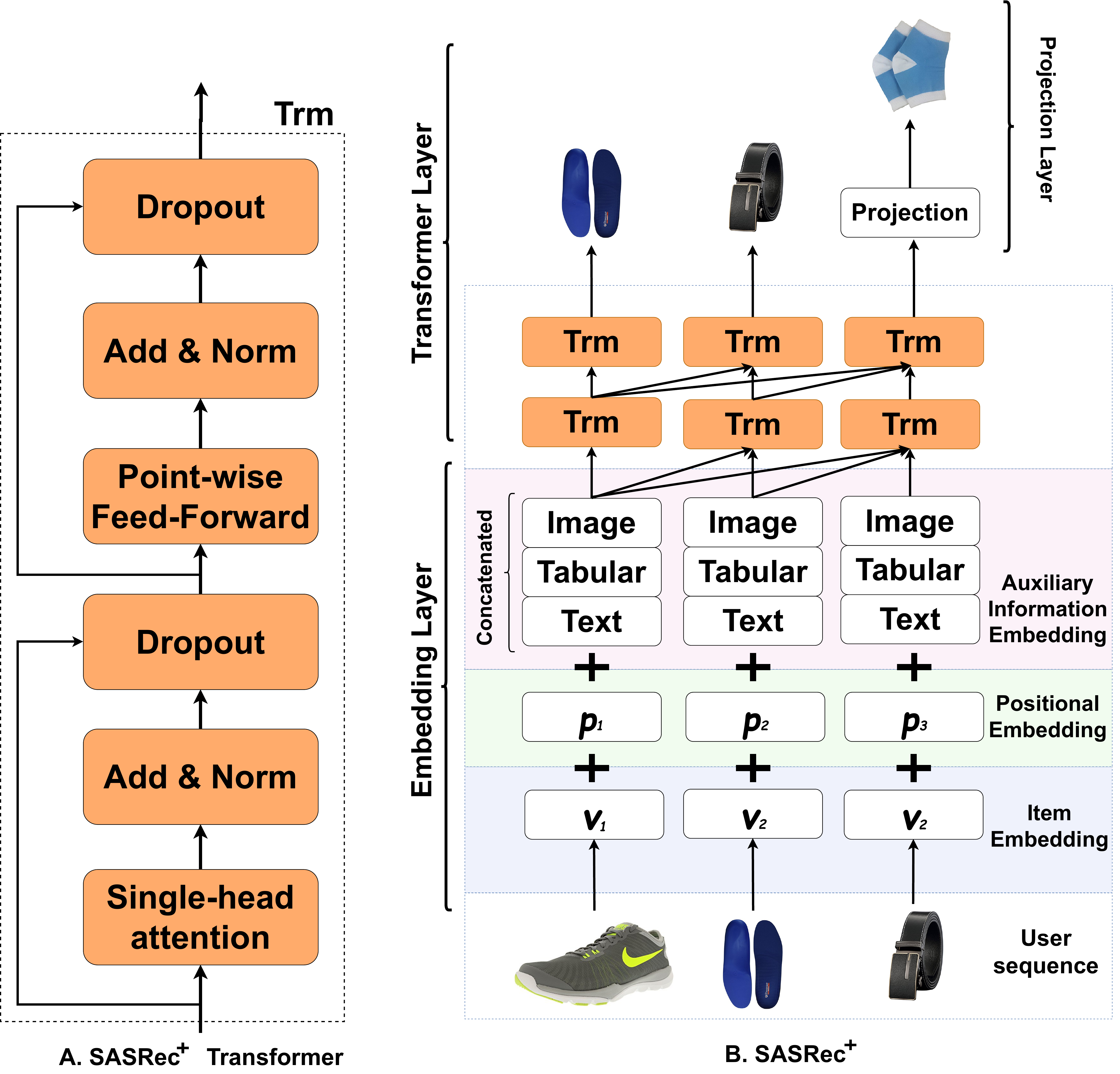}}
\caption{The SASRec$^+$ architecture with the inclusion of multi-modal auxiliary information embedding}
\label{figure:Models}
\end{figure}

\section{Experimental Methodology}

\subsection{Problem Statement}
The problem addressed in this paper is formulated as follows. Given a set of users $\mathcal{U} = \{u_1, u_2, \dots, u_{|\mathcal{U}|}\}$, a set of items $\mathcal{V} = \{v_1, v_2, \dots, v_{|\mathcal{V}|}\}$, each item's auxiliary information embedding $\mathcal{K} = \{\mathbf{k}_1, \mathbf{k}_2, \dots, \mathbf{k}_{|\mathcal{V}|}\}$ and the sequential item data $\mathcal{S}_u = \{v_1^u, v_2^u, \dots, v_{n_u}^u\}$, predict the next item $v_{n_u + 1}^u$ that is most preferred by user $u$, where $n_u$ is the number of items rated by user $u$, $v_t^u$ ($1 \leq t \leq n_u$) represents the relative time $t$ when user $u$ interacted with item $v$, and $\mathbf{k}_v$ is a joint vector embedding of text, tabular and image data for item $v$.


\subsection{Multi-Modal Auxiliary Information}

Auxiliary information may exist as tabular or unstructured data. Some research is focused on tabular data, usually categorical data~\cite{fischer2020integrating, vasile2016meta, arunmetatransformer4rec}. Practitioners transform the categorical data by one-hot encoding the embeddings. In contrast, this study uses vector representations transformed into embeddings from tabular and unstructured data.  

The study uses data2vec to extract embeddings from unstructured data~\cite{baevski2022data2vec}. Baevski~\textit{et al.} proposed a novel framework that uses the same learning method for speech, natural language and images by predicting latent representations of the input data. Mainly, data2vec is a general self-supervised learning method that predicts the latent representations of the full input data using a standard transformer architecture. 

\textbf{dat2vec Text:}
The study builds on the pre-trained data2vec text model from HuggingFace~\cite{huggingfacedata2vec}. Data2vec was pre-trained on five datasets consisting of texts from books, Wikipedia and news articles. We begin by tokenizing the text data using a byte-pair encoding~\cite{sennrich2015neural}. Then we pass the array of tokens of size 196 to the data2vec text model and use the last hidden state as its embeddings. A number greater than 196 results in overly large parameters for training the transformers. Whereas a number much smaller than 196 may result in information loss.

\textbf{dat2vec Image:}
The data2vec vision model was pre-trained on the ImageNet-1k fashion dataset. We resize each image in our training and testing datasets into 224x224 RGB images. Furthermore, we follow the same feature extraction process as BEiT and normalize the images~\cite{bao2021beit}. Similarly, we retain the last hidden state, of size 196, as the image embedding.

\textbf{XGBoost Tabular:}
The study selects a gradient-boosting approach over deep learning methods for tabular data. Deep learning methods do not outperform gradient-boosted tree ensembles for classification and regression problems with tabular data~\cite{shwartz2022tabular}. Hence, we extract embeddings from tabular data using XGBoost~\cite{chen2015xgboost}. We begin by one-hot encoding all categorical features in the table and ensure that only numerical types are present within the dataset. We train XGBoost on 70\% of the data and set the training target variable as the user ratings. We perform 4-fold cross-validation to select the model with the least Mean Absolute Error.

Finally, we construct a joint embedding $\mathcal{k}_v$ of text, tabular and image data for item $v$. The embedding is either the concatenations or the summation of the text, tabular and image representations. The paper uses the concatenation approach to compare our proposed approach to baselines.

\subsection{Embedding Layer}
To integrate auxiliary information consisting of both tabular and unstructured data into a transformer architecture, we take an embedding approach inspired by KeBert4Rec~\cite{fischer2020integrating}. The embedding layer for KeBert4Rec consists of three different learned embeddings: (i) an item identifier embedding $E_{\mathcal{V}} \in \mathbb{R}^{|\mathcal{V}|\times d}$, (ii) a positional embedding $E_{\mathcal{P}} \in \mathbb{R}^{|N|\times d}$ and (iii) an auxiliary information embedding $\mathcal{K}$. The positional embedding encodes the positions of the items within a user's sequential item data $\mathcal{S}_u$. $N$ is the configurable maximum input sequence length. The auxiliary information embedding encodes categorical data into a multi-hot encoded vector. The vector is scaled to the embedding size $d$ using a linear layer. For each item in the sequence $\mathcal{S}_u$, we have the item embedding $e_{v_t} = E_{\mathcal{V}}v_t$, the positional embedding $p_{v_t} = E_{\mathcal{P}}t$ and auxiliary information embedding $\mathbf{k}_{v_t}$. Where $t$ is the time at which the user consumed the item. The sum $h_t^0 = e_{v_t} + p_{v_t} + k_{v_t}$ is used as the input into the transformer.

The critical difference in our implementation, compared to KeBert4Rec, is that we do not use a multi-hot encoding of categorical data. Instead, we make use of all multi-modal auxiliary information available and transform it into an embedding of continuous values.
The study investigates whether concatenated or summation (point-wise addition of vector items) produces the best embeddings. 

Finally, we sum the item, positional and auxiliary information embeddings and pass it as input to the transformer layer. Fig.~\ref{figure:Models}B depicts the embedding layer for the unidirectional model SASRec. We use the same embedding layer for the bidirectional model BERT4Rec.

\subsection{Transformers}
The study adopts uni- and bi-directional transformer architecture. The state-of-the-art uni- and bi-directional transformers for sequential recommendation are SASRec and BERT4Rec, respectively~\cite{kang2018self, sun2019bert4rec}. Given an input sequence, both transformers iteratively compute hidden representations $h_i^l$ at each layer $l$ for each position $i$, simultaneously. The scaled dot-product attention is~\cite{vaswani2017attention}:
\begin{equation}
    \mathrm{Attention(\mathbf{Q, K, V})} = \mathrm{softmax} \left( \frac{\mathbf{QK}^T}{M} \mathbf{V} \right) \label{equation:Attention}
\end{equation}
where \textbf{Q} represents the queries, \textbf{K} the keys and \textbf{V} the values
(each row represents an item). $M$ is a scale factor used to avoid overly large inner product values.

\subsubsection{Self-Attention} SASRec uses a single-head self-attention mechanism, whereas BERT4Rec uses a multi-head self-attention mechanism. SASRec attends to information from left to right, whereas BERT4Rec attends to information from different representation subspaces at different positions.
Take the input embedding \textbf{H}$^l$; we convert it into three matrices through linear projections, for SASRec:
\begin{equation}
    \mathrm{SA}(\textbf{H}^l) = \mathrm{Attention}(\textbf{H}^l \mathbf{W}^\mathrm{Q}, \mathbf{H}^l \mathbf{W}^\mathrm{K}, \mathbf{H}^l \mathbf{W}^\mathrm{V}) \label{equation:SelfAttention}
\end{equation}
and $h$ subspaces for BERT4Rec
\begin{equation}
    \mathrm{MH}(\mathbf{H}^l) = [\mathrm{head}_1, \mathrm{head}_2, \dots, \mathrm{head}_h] \mathbf{W}^O \label{equation:Multi-Head}
\end{equation}
\begin{equation}
\mathrm{head}_i (\mathbf{H}^l) = \mathrm{Attention}(\mathbf{H}^l \mathbf{W}_i^\mathrm{Q}, \mathbf{H}^l \mathbf{W}_i^\mathrm{K}, \mathbf{H}^l \mathbf{W}_i^\mathrm{V}) \label{equation:Head}
\end{equation}
where $\mathbf{W}_i^\mathrm{Q} \in \mathbb{R}^{d \times d/h}$, $\mathbf{W}_i^\mathrm{K} \in \mathbb{R}^{d \times d/h}$ and $\mathbf{W}_i^\mathrm{V} \in \mathbb{R}^{d \times d/h}$ are the projection matrices with learnable parameters. $h = 1$ for SASRec.

\subsubsection{Feed-Forward Network} 
To endow the model with non-linearity and consider interactions between different dimensions, SASRec adopts a point-wise Feed-Forward Network 
\begin{equation}
    \mathrm{FFN}_{S}(\mathbf{h}_i^l) = \textnormal{ReLU} (\mathbf{h}_i^l \mathbf{W}^{(1)} + \mathbf{B}^{(1)}) \mathbf{W}^{(2)} + \mathbf{B}^{(2)} \label{equation:FFN_SASRec}
\end{equation}
and Bert4Rec adopts a position-wise Feed-Forward Network
\begin{equation}
    \mathrm{PFFN}_{B}(\mathbf{h}_i^l) = [\mathrm{FFN}_{B}(\mathbf{h}_1^l)^T, \mathrm{FFN}_{B}(\mathbf{h}_2^l)^T, \dots, \mathrm{FFN}_{B}(\mathbf{h}_t^l)^T] \label{equation:FFN}
\end{equation}
\begin{equation}
    \mathrm{FFN}_{B}(\mathbf{h}_i^l) = \textnormal{GELU} (\mathbf{h}_i^l \mathbf{W}^{(1)} + \mathbf{B}^{(1)}) \mathbf{W}^{(2)} + \mathbf{B}^{(2)} \label{equation:FFN_BERT4Rec}
\end{equation}
with a Gaussian Error Linear Unit (GELU) activation. $\mathbf{W}^{(1)}, \mathbf{W}^{(2)}  \in \mathbb{R}^{d \times d}$ and $\mathbf{d}^{(1)}, \mathbf{d}^{(2)}  \in \mathbb{R}^{d}$ are learnable parameters.

Fig.~\ref{figure:Models} depicts the structure of the proposed model $\mathrm{SASRec}^+$, which is a slight modification of SASRec for multi-model data.

\textbf{Proposed Method:} 
The study adopts a similar design to the KeBERT4Rec transformer architecture~\cite{fischer2020integrating}. KeBERT4Rec integrates auxiliary information into the bidirectional transformer BERT4Rec~\cite{sun2019bert4rec}, along with the sequential item data. Denote the sequential item data for user $u$ as $\mathcal{S}_u = \{v_1^u, v_2^u, ..., v_{n_u}^u\}$, where $n_u$ is the number of items rated by user $u$ and $v_t^u$ ($1 \leq t \leq n_u$) represents the relative time $t$ when user $u$ interacted with item $v$. We consider for each item $v \in \mathcal{V}$ an embedding of structured and unstructured data $\mathbf{k}_v$. Given the history $\mathcal{S}_u$ of user $u$ and the auxiliary information $\mathcal{K}_u = \{\mathbf{k}_{v_t^u}, 0 \leq t \leq n_u\}$ that corresponds to $\mathcal{S}_u$, the transformer must predict $v_{n_u + 1}^u$ from the sequence of user's interaction and its corresponding auxiliary information. It can be formalized as modelling the probability over all possible items for user $u$ at relative timestamp $n_u + 1$:
\begin{equation}
    p(v_{n_u + 1}^{u}) = (v \vert \mathcal{S}_u, \mathcal{K}_{u}) \label{equation:RecommendationProbaility}
\end{equation}
We refer to the uni- and bi-directional model that includes multi-model data as SASRec$^+$ and BERT4Rec$^+$.

\subsection{Datasets}

The study explores the Amazon Fashion dataset and ML-20M. The Amazon Fashion dataset contains a moderate amount of interactions with a relatively sparse interaction matrix~\cite{ni2019justifying}. ML-20M is a benchmark dataset widely used in the literature, and it has more users than items~\cite{harper2015movielens}. However, the interactions in ML-20M  are usually more evenly distributed across users than items. Each dataset consists of tabular, text and image data and the time a user consumed an item.

\begin{figure}[ht]
\begin{subfigure}[b]{\textwidth}
\centering
\includegraphics[width=0.8\columnwidth]{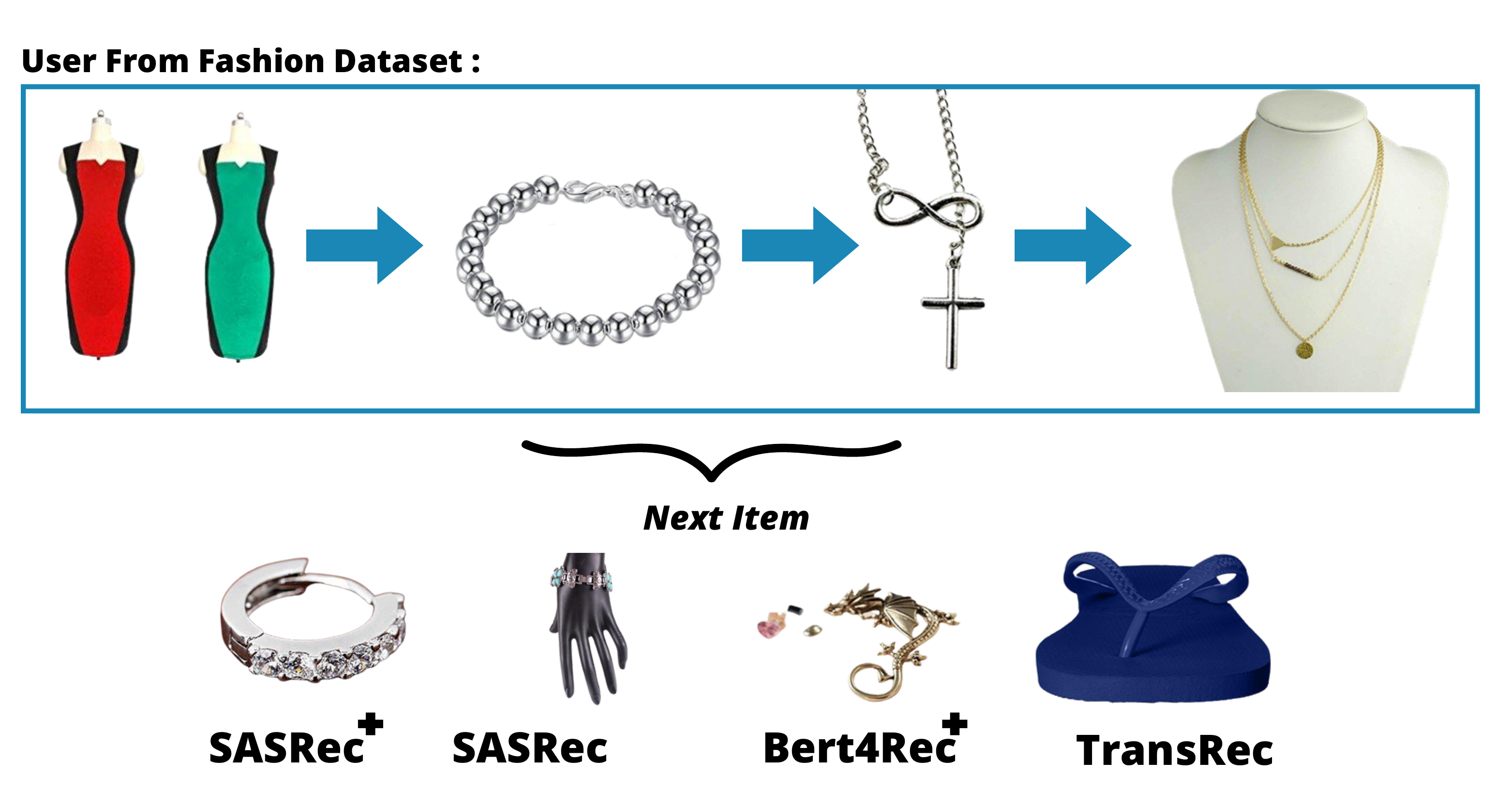}
\end{subfigure}
\begin{subfigure}[b]{\textwidth}
\centering
\includegraphics[width=0.8\columnwidth]{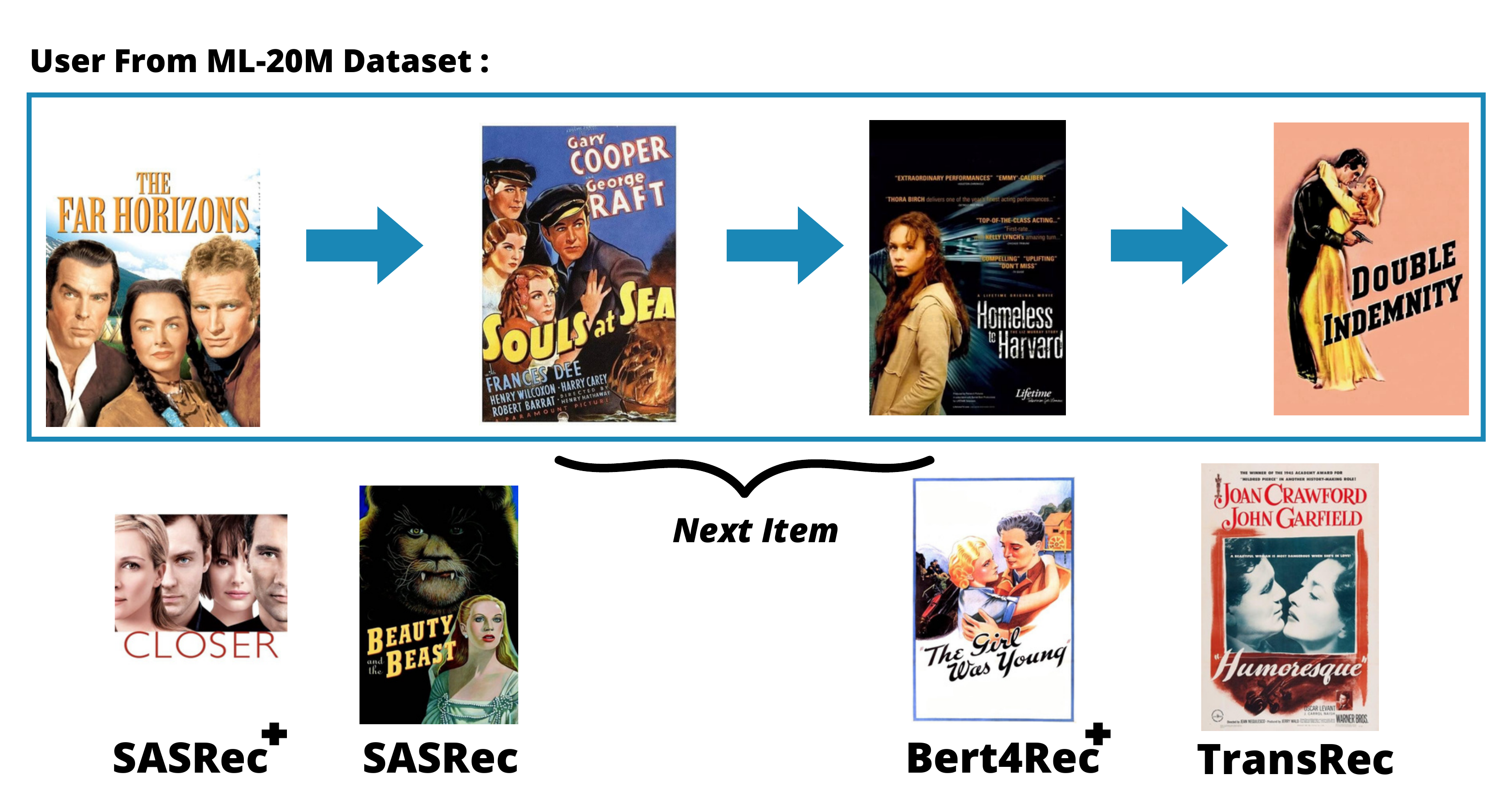}
\end{subfigure}
\caption{A sample of two users' historical consumption of items and the predicted next item by each model}
\label{figure:sample}
\end{figure}

\subsection{Baselines}

The study uses the following three groups of recommendation baselines. 
\begin{enumerate}
    \item General recommendation methods that only consider user feedback without modelling sequential behaviour. The methods considered in this group consist of the following:
    \begin{itemize}
        \item \textbf{PopRec} which ranks items according to their popularity based on the number of interactions~\cite{steck2011item}; 
        \item \textbf{Bayesian Personalized Ranking (BPR)} optimizes matrix factorisation with implicit feedback using pairwise Bayesian Personalized Ranking~\cite{rendle2012bpr}.
    \end{itemize}
    \item Sequential recommendation methods based on first-order Markov chains. \textbf{Translation-Based Recommendation (TransRec)} is a state-of-the-art approach that models each user as a translation vector to capture item transitions~\cite{he2017translation}.
    \item Selected state-of-the-art deep learning-based sequential recommendation systems, namely, \textbf{SASRec}, \textbf{BERT4Rec} and \textbf{KeBERT4Rec}.    
\end{enumerate}
Fig~\ref{figure:sample} gives an example of a user from each dataset, and their sequential consumption of items. The diagram also depicts the predicted next item by each sequential recommendation model.

\subsection{Evaluation}

Following previous research, we applied the \textit{leave-one-out} strategy to recommend the next item~\cite{kang2018self, sun2019bert4rec, zhou2020s3}. For each user interaction sequence, we used the last item as the test data and the item prior to it as the validation data. The remaining items are used for training. The entire item set is usually too large to rank all candidate items. Hence, we randomly selected 100 negative items (the user has not interacted with) and rank these items together with the test item (ground truth or positive item)~\cite{kang2018self, sun2019bert4rec}. Items are ranked based on their respective predicted preference scores. Subsequently, given a user, the recommendation task is to identify which item is the ground truth next item amongst the 101 items (i.e., 100 negative items plus the ground truth item). 

We evaluated the proposed method against the baselines using three performance metrics. These three performance metrics assess the relevancy of the predicted items concerning the user. 
Hit Rate @N ($\mathrm{HR@N}$) measures the fraction of users for which a query of size $N$ included the correct item. A large query size returns a higher hit ratio because there is an increased chance of the query containing the item. 

In addition, we chose two position-aware ranking-based performance metrics: 
Normalized Discounted Cumulative Gain ($\mathrm{NDCG@N}$) assigns higher weights to the most relevant item ranked at the top of the query. Simultaneously, if the user's most preferred item is ranked low by the model, the NDCG algorithm penalises the model by assigning lower weights. A large query size returns a higher $\mathrm{NDCG}$ score since the most preferred item now sits above more irrelevant items in the query. 

Lastly, Mean Average Precision ($\mathrm{MAP}$) consists of two parts; (i) Precision@N measures the fraction of items most preferred by the user in a query of size $N$ (ii) average Precision@N is the sum of Precision@N for different values of $N$ divided by the total number of preferred items in the query. Consequently, $\mathrm{MAP}$ is the average of average Precision@N over all the queries in the entire dataset.

This study selects the $\mathrm{NDCG@10}$ score as the key metric. Our objective is to ensure that the most preferred item is ranked highest within a query of 10 randomly selected items. $\mathrm{NDCG@10}$ is an accuracy-based metric that emphasises the correct positioning of items in a query. Hence, a model that achieves the highest $\mathrm{NDCG@10}$ score better conceptualises the user's behaviour and preferences.

\begin{table}[htb!]
\caption{Performance of each recommendation model. Top two models in grey. Best model in bold$^a$.}\label{table:Results}
\centering
\resizebox{\textwidth}{!}{%
\begin{tabular}{lrrrrrrrrr}
\makecell[lb]{
\textbf{Dataset}\\
\quad \textbf{Metrics}} & 
\textbf{\makecell[br]{Pop-\\Rec}} &
\textbf{BPR} & 
\textbf{\makecell[br]{Trans-\\Rec}} & 
\textbf{\makecell[br]{BERT-\\4Rec}} & 
\textbf{\makecell[br]{Ke-\\BERT-\\4Rec}} & 
\textbf{\makecell[br]{SAS-\\Rec}} & 
\textbf{\makecell[br]{BERT-\\4Rec$^+$}} & 
\textbf{\makecell[br]{SAS-\\Rec$^+$}}& 
\textbf{Improve$^b$} \\
\toprule
         \textbf{ML-20M} &&&&&&&&& \\        
         \quad HR@1    & .014 & .202 & .217 & .227 & .229 & \rf{.429} & .229      & \bf{.431} & 0.4 \% \\
         \quad HR@5    & .054 & .489 & .563 & .569 & .568 & \rf{.802} & .570      & \bf{.818} & 2.0 \% \\
         \quad HR@10   & .081 & .630 & .755 & .740 & .746 & \rf{.904} & .746      & \bf{.920} & 1.7 \% \\
         \quad NDCG@5  & .034 & .350 & .394 & .403 & .404 & \rf{.629} & .405      & \bf{.638} & 1.4 \% \\
         \quad NDCG@10 & .043 & .396 & .457 & .445 & .461 & \rf{.679} & .462      & \bf{.706} & 4.0 \% \\
         \quad MAP     & .046 & .338 & .379 & .387 & .387 & .387      & \rf{.388} & \bf{.597} &53.9 \% \\
 \midrule
         \textbf{Fashion} &&&&&&&&& \\        
         \quad HR@1    & .001 & .049 & .057 & .049 & .069 & \rf{.437} & .075 & \bf{.525} & 16.9 \% \\
         \quad HR@5    & .001 & .208 & .214 & .199 & .244 & \rf{.760} & .244 & \bf{.883} & 16.3 \% \\
         \quad HR@10   & .020 & .229 & .266 & .284 & .266 & \rf{.870} & .267 & \bf{.955} &  9.7 \% \\
         \quad NDCG@5  & .001 & .130 & .133 & .122 & .162 & \rf{.608} & .166 & \bf{.720} & 18.4 \% \\
         \quad NDCG@10 & .007 & .136 & .151 & .161 & .169 & \rf{.643} & .176 & \bf{.713} & 11.0 \% \\
         \quad MAP     & .015 & .114 & .126 & .120 & .149 & \rf{.580} & .154 & \bf{.677} & 16.8 \% \\
\bottomrule
\multicolumn{10}{l}{$^a$ \textbf{BERT4Rec$^+$} and \textbf{SASRec$^+$} are the proposed methods.} \\
\multicolumn{10}{l}{$^b$ percentage improvement achieved by \textbf{SASRec$^+$} against the best baseline.}
\end{tabular}}
\end{table}

\section{Results}

Table~\ref{table:Results} presents the performance of all models on the two datasets. BERT4Rec$^+$ and SASRec$^+$ are our proposed implementations. The $+$ refers to the inclusion of multi-modal auxiliary information. At first glance, SASRec$^+$ scores the highest against each baseline across all metrics. With SASRec achieving the second-highest performance. The results show that a unidirectional transformer architecture is far superior in modelling context data. In addition, we observe that BERT4Rec only experiences marginal gains from including multi-modal data. The NDCG@10 score only improves by 0.04\% and 0.09\% for the ML-20M and Fashion dataset, respectively. As expected, non-context-aware models, PopRec and BPR, perform worse than the other models across all metrics. Proving that information about a user's sequential consumption of items is essential in understanding a user's preference. 

Table~\ref{table:Results} shows that the highest performance gains were achieved on the MAP score by SASRec$^+$.
Precisely, our model is 54\% and 17\% more capable of uncovering the most preferred items above a model without auxiliary information, for the ML-20M and Fashion dataset, respectively. Therefore, multi-modal auxiliary information adds further context that aids in understanding the item's characteristics.
\begin{table}[htb!]
\caption{Paired t-test of NDCG@10 scores with and without multi-modal auxiliary information}\label{table:Significance}
\centering
\resizebox{\textwidth}{!}{%
\begin{tabular}{llrrrr}
\textbf{\makecell[bl]{Dataset}} & 
&
\textbf{\makecell[br]{BERT4Rec}} & 
\textbf{\makecell[br]{BERT4Rec$^+$}} & 
\textbf{\makecell[br]{SASRec}} & 
\textbf{\makecell[br]{SASRec$^+$}} \\
\midrule
& Mean & 	.4450 & .4615 & .6787 & .7060 \\
& Standard deviation & .0009 & .0004 & .0187 & .0026 \\
\textbf{ML-20M} & Observations & & 10 &  & 10 \\
& degrees of freedom  & & 9 & & 9 \\
& t statistic & & 64.2321 & & 4.0780 \\
& p-value & & 2.7e-13 & & .0028 \\
\cmidrule{2-6}
& Mean & 	.1605 & .1757 & .6428 & .7134 \\
& Standard deviation & .0128 & .0089 & .0117 & .0326 \\
\textbf{Fashion} & Observations & & 10 & & 10 \\
& degrees of freedom  & & 9 & & 9 \\
& t statistic & & 3.7576 & & 5.5814 \\
& p-value & & .0045 & & .0003 \\
\bottomrule
\end{tabular}
}
\end{table}
Notably, all three BERT4Rec implementations perform only slightly better than TransRec and substantially below all two SASRec variants. Presumably, the bidirectional structure of BERT4Rec is trained on both past and future items consumed by a user. Whereas a unidirectional model like SASRec is only trained on the past values consumed by a user. The poor performance of the bidirectional transformer proves that a user does not consume an item based on future consumption patterns. However, for each uni- and bi-directional model, including auxiliary information, adds to the performance of each model. Table~\ref{table:Significance} reports on the statistical significance of adding auxiliary information into the two transformer architectures. We observe a p-value less than 0.01 for BERT4Rec$^+$ and SASRec$^+$. Hence, multi-modal auxiliary information provides a better context of a user and item's relationship. 

\subsection{Ablation Study}

\begin{table}[htb!]
\caption{Ablation analysis (NDCG@10) on two datasets}\label{table:Ablation}
\centering
\resizebox{\textwidth}{!}{%
\begin{tabular}{lccccccc}

\quad \textbf{Dataset} & \multicolumn{2}{c}{\textbf{ML-20M}}   & \multicolumn{3}{c}{\textbf{Fashion}}\\

 \textbf{Transformers} & \textbf{BERT4Rec$^+$} & \textbf{SASRec$^+$} &  & \textbf{BERT4Rec$^+$} & \textbf{SASRec$^+$}\\
\midrule
    \quad (1) Text & .46181 & .70260 & & .18384 & \bf{.72807} \\
    \quad (2) Image & \rf{.46219} & .68590 & & \bf{.20737} & .58797 \\
    \quad (3) Tabular & .46182 & .68815 & & .19242 & .59929 \\
    \cmidrule{2-6}
\textbf{Concatenated} \\
    \quad (4) Text + Image + Tabular & .46146 & .70600 & & .17567 & .71342\\
    \quad (5) w/o Tabular & .46203 & .70676 & & .18828 & .69398\\
    \quad (6) w/o Image & .46198 & \rf{.70837} & & .17734 & \rf{.72643}\\
    \quad (7) w/o Text & .46182 & \bf{.71004} & & .15517 & .62355\\
   \cmidrule{2-6}
\textbf{Summation} \\
    \quad (8) Text + Image + Tabular & .46218 & .67628 & & .19266 & .43453 \\
    \quad (9) w/o Tabular & .46211 & .68225 & & .19236 & .53307 \\
    \quad (10) w/o Image & .46198 & .65881 & & \rf{.19469} & .65882 \\
    \quad (11) w/o Text & \bf{.46226} & .68386 & & .17060 & .41108 \\
\bottomrule
\multicolumn{6}{l}{Top two best performing variants along the column are in grey. Best model in Bold}
\end{tabular}
}
\end{table}

We performed an ablation study of the impact of each modality on both the uni- and bi-directional transformer. We recorded the results of a combined embedding of modalities and investigated whether a concatenated or summation approach works best. Table~\ref{table:Ablation} reports the NDCG@10 scores achieved for each variant over SASRec$^+$ and BERT4Rec$^+$. The essential characteristic of the ML-20M dataset is that it is a large dataset with long item sequence vectors, an average length of 65. In contrast, the Fashion dataset is relatively sparse with concise item sequence vectors, an average length of six.

We observe that short sequence datasets benefit most from a single modality embedding (see rows (1) - (3) in Table~\ref{table:Ablation}), whereas longer sequence datasets produce the best results from a joint embedding. Long sequence datasets produce more complex relationships between different users' behaviour. Hence, we can uncover similarities between items within the embedding space by extending our knowledge of the item. Often the bi-directional transformer yields the highest results from an embedding that includes image data (see rows (2), (6) and (10)), whereas the uni-directional model requires text data(see rows (1) and (11)). In addition, a concatenated joint embedding works best for a uni-directional model, whereas a bi-directional model can exploit the summation approach (see rows (4) and (8), respectively). Bi-directional models worsen with longer text descriptions or concatenated joint embeddings. Similarly, uni-directional models worsen with image data and the summation of embeddings. Presumably, both models experience some information loss. The multi-head self-attention mechanism deteriorates from longer and more detailed data formats, such as text descriptions and concatenated embeddings. The uni-directional model deteriorates from less descriptive data formats such as images and summation of embeddings.

\begin{figure}[htp]
  \centering
  \begin{tabular}{cc}
  \includegraphics[width=0.5\linewidth]{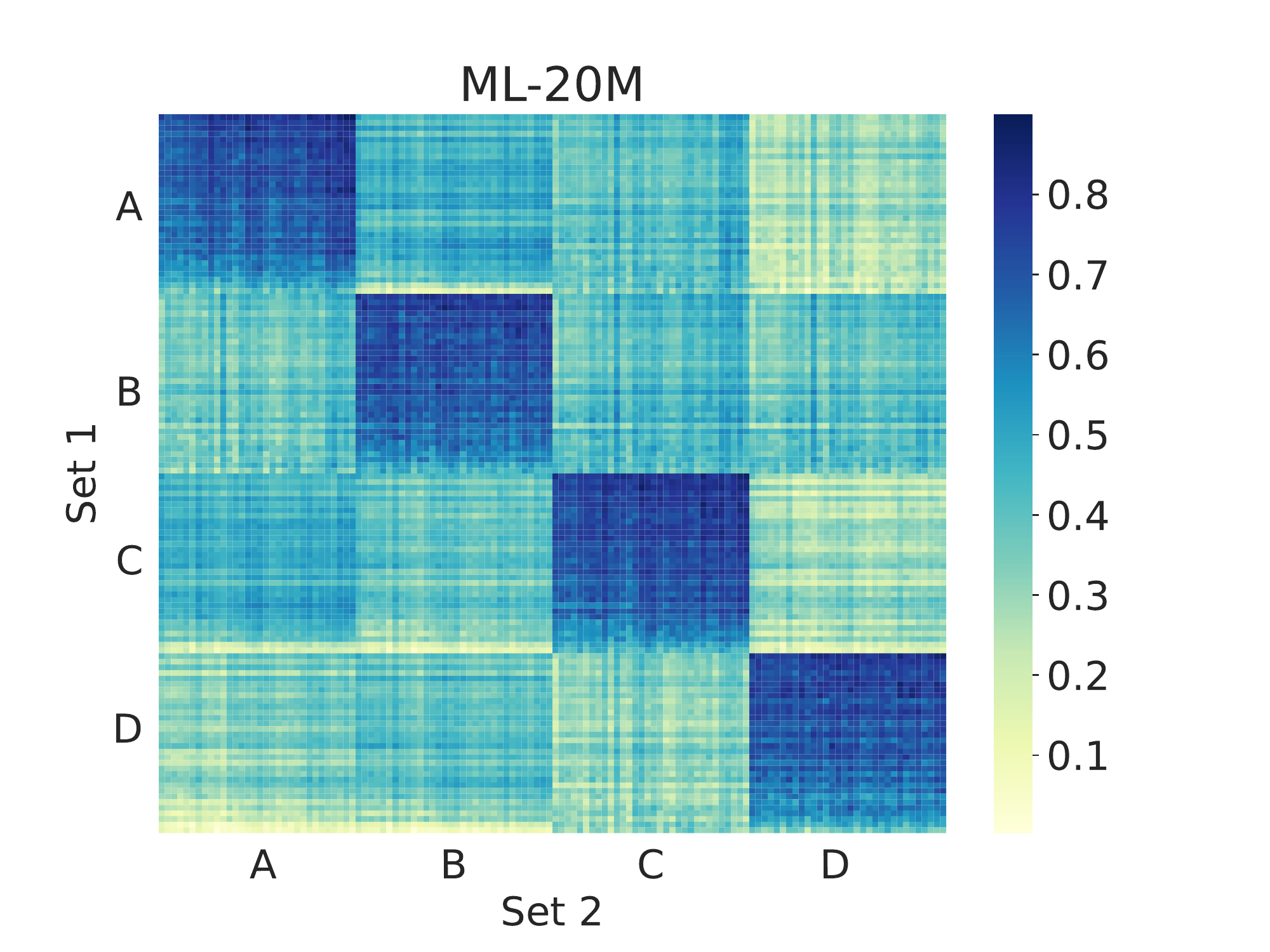} & 
   \includegraphics[width=0.5\linewidth]{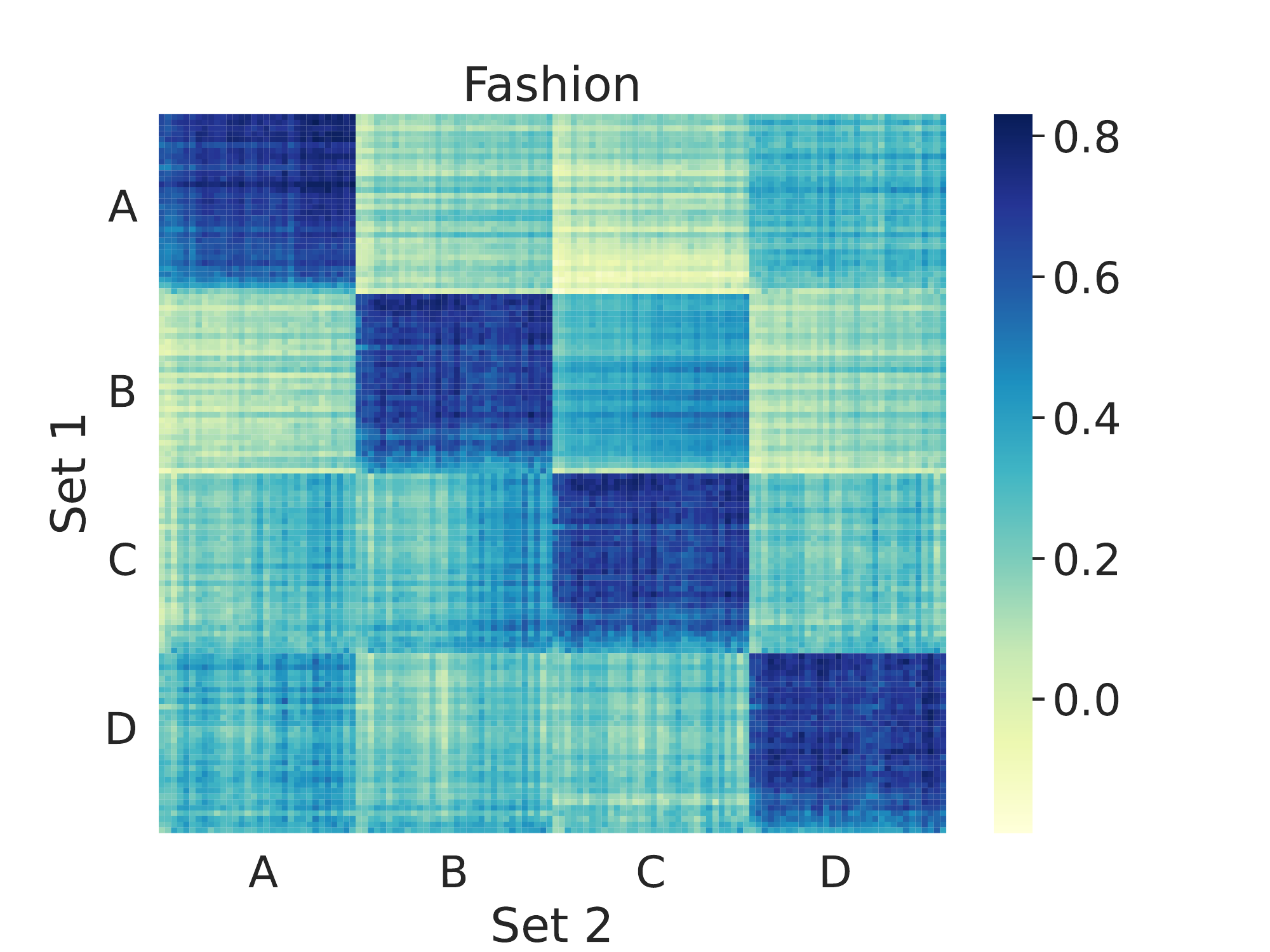}\\ 
  \end{tabular} 
  \caption{Heatmaps of the similarity scores between two sets of user's average attention weights}\label{figure:Heatmap}
\end{figure}

\subsection{Visualizing Attention Weights}

Finally, we show that our model is user context-aware. Our objective is to demonstrate that we can learn the context of each user by looking at sequential and multi-modal data. We do not want a scenario where a recommendation is being made by clustering similar items. Instead, we want to group users by their behaviour and recommend items from other users that share similar behaviour. The paper investigates if our model is user context-aware by visualizing the attention weights from the best performing model, i.e. SASRec$^+$. 

First, we use the K-means algorithm to group users into clusters according to their behaviour. Then we use the naive elbow method to select the optimal number of clusters that separates the users at a more significant level. 
We had seven and five optimal clusters for the ML-20M and Fashion datasets, respectively. We selected four clusters for visualization purposes. Each user sequence vector belongs to one of the four clusters, denoted by A, B, C or D for brevity. Then, we randomly select two disjoint sets from each cluster. Each set contains 100 users; therefore, we have 200 users per cluster. We pass each of the 200 user's sequence vectors and auxiliary information into SASRec$^+$ and obtain their average attention weights.

Lastly, we calculate the cosine similarity of the average attention weights between sets one and two for each cluster. Our goal is to observe the highest similarity between sets one and two that belong to the same cluster. Fig.~\ref{figure:Heatmap} depicts two heatmaps of the similarity scores of the average attention weights between the two sets, for ML-20M and Fashion datasets, respectively. Both heatmaps are approximately block diagonal matrices, meaning that SASRec$^+$ can identify users that share similar behaviour and tend to assign larger weights between them. This adds evidence to our model being user context-aware.

\section{Conclusion}

This paper aimed to design a multi-modal recommendation system with auxiliary information. We used data2vec to obtain embeddings from multi-modal data. We combined these embeddings and then passed them to a uni- or bi-directional transformer to model the behaviour of sequential consumption of items for each user.

The study found that auxiliary information improves the prediction results. Notably, a unidirectional transformer benefits most from multi-modal auxiliary information than a bidirectional one. As a result, we conclude that multi-model data adds further context to a user's behaviour. Hence, a multi-modal recommendation system benefits more from a model that utilizes an item's multi-modal data together in the embedding space.

This study used a general learning framework called data2vec to obtain embeddings. However, we did not assess the quality of the image and text embeddings obtained from data2vec. Though from the ablation study, we observed inconsistent results when combining different modalities, we conjecture that the embeddings obtained from the text and image data, belonging to the same item, are similar. We encourage future research to investigate the similarity of embeddings belonging to the same item but exist in different modalities.

%
%
%
\bibliographystyle{splncs04}
\bibliography{main}

\end{document}